\documentclass[12pt,a4,cite]{article}

\newcommand{\ba}{\begin{eqnarray}}
\newcommand{\ea}{\end{eqnarray}}
\newcommand{\be}{\begin{equation}}
\newcommand{\ee}{\end{equation}}
\newcommand{\dis}{\displaystyle}

\usepackage{axodraw}
\usepackage{epsfig}

\newcommand{\noi}{\noindent}
\newcommand{\nn}{\nonumber}
\newcommand{\tr}{\mbox{\rm tr}}

\newcommand{\ksls}{\not \! k}
\newcommand{\psls}{\not \! p}

\newcommand{\pslsh}{\not \! p}

\begin{document}

\begin{titlepage}
\begin{flushright}
CAFPE-120/09\\
UG-FT-250/09
\end{flushright}
\vspace{2cm}
\begin{center}

{\large\bf  The Hadronic Light-by-Light Contribution to Muon
$g-2$: a Short Review
\footnote{Invited talk at ``Photon09, 
International Conference on 
the Structure and the Interactions of the Photon'', 
May 11-15  2009, DESY, Hamburg, Germany.}}\\
\vfill
{\bf  Joaquim Prades}\\[0.5cm]
CAFPE and Departamento de
 F\'{\i}sica Te\'orica y del Cosmos, Universidad de Granada, 
Campus de Fuente Nueva, E-18002 Granada, Spain.\\[0.5cm]

\end{center}
\vfill

\begin{abstract}
I review the recent calculations and current status of the hadronic 
light-by-light
scattering contribution to muon g-2. In particular,  I discuss the main results
 obtained in  a recent work together with Eduardo de Rafael and Arkady Vainshtein
where we came to the estimate 
$a^{\rm HLbL}_\mu = (10.5 \pm 2.6) \times 10^{-10}$.
How the two-photon physics program of low energy facilities can help
to reduce the present model dependence is also emphasized.
\end{abstract}

\end{titlepage}

\section{Introduction}

\begin{figure}[hbt]
\begin{center}
\epsfig{file=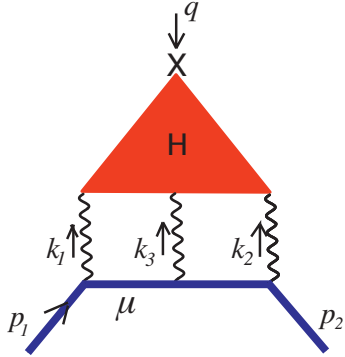,width=4.5cm}
\end{center}
\caption{Hadronic light-by-light scattering contribution.}
\label{fig:1}      
\end{figure}

One momenta configuration out of the six possible ones
contributing to the  hadronic light-by-light to muon g-2 is depicted
in Fig. \ref{fig:1} and described by the  vertex function 
\ba
\label{Mlbl}
\dis{\Gamma^\mu} (p_2,p_1)
&=&  - e^6  
\int {{\rm d}^4 k_1 \over (2\pi )^4}
\int {{\rm d}^4 k_2\over (2\pi )^4}  
{\Pi^{\mu\nu\rho\sigma} (q,k_1,k_2,k_3) 
\over k_1^2\, k_3^2 \, k_3^2} \nonumber \\ &\times&  
\gamma_\nu (\not{\! p}_2+\not{\! k}_2-m )^{-1} 
\gamma_\rho (\not{\! p}_1-\not{\! k}_1-m )^{-1} \gamma_\sigma \, 
 \nonumber \\ 
\ea
where $q \to 0 $ is the momentum of the
photon that couples to the external magnetic source,
$q=p_2-p_1=-k_1-k_2-k_3$ and $m$ is the muon mass. 

 The dominant contribution to the hadronic four-point  function 
\ba
\label{four}
\Pi^{\rho\nu\alpha\beta}(q,k_1,k_3,k_2)&=& \nonumber \\
i^3 \int {\rm d}^4 x \int {\rm d}^4 y
\int {\rm d}^4 z \, \, {\rm e}^{i (-k_1 \cdot x + k_3 \cdot y + 
k_2 \cdot z)}
 \,   \langle 0 | T \left[  V^\mu(0) V^\nu(x) V^\rho(y) V^\sigma(z)
\right] |0\rangle && 
\ea
comes from the three light quark 
$(q = u,d,s)$ components in the electromagnetic current
$V^\mu(x)=\left[ \overline q \, \widehat Q \, \gamma^\mu \, q \right](x)$
where $\widehat Q \equiv {\rm diag} (2, -1, -1)/3$ 
denotes the quark electric charge matrix. 
We are interested in the limit $q \to 0$ where
current conservation implies
\ba
\dis{\Gamma^\mu} (p_2,p_1) =
- \frac{a^{\rm HLbL}}{4 m} \, 
\left[\gamma^\mu, \gamma^\nu \right] \, q_\nu \, .
\ea

Therefore, the muon anomaly can then be extracted as
\ba
\label{aH}
a^{\rm HLbL}\!\!\!\!&=&\!\!\!\!\frac{e^{6}}{48m}\int\!\frac{d^4 k_1}{(2\pi)^4}\!\!\int\frac{d^4 k_2}{(2\pi)^4}\frac{1}{k_1^2 k_2^2 k_3^2}
\left[\frac{\partial}{\partial q^{\mu}}
\Pi^{\lambda\nu\rho\sigma}(q,k_1, k_3, k_2)\right]_{q=0}  \nonumber\\[1mm] 
 & &\times \,\tr\left\{
        (\psls + m)[\gamma_{\mu},\gamma_{\lambda}](\psls +m) \gamma_{\nu}(\pslsh+\ksls_2-m)^{-1}\gamma_{\rho}(\pslsh -\ksls_1-m)^{-1}\gamma_{\sigma}\right\}\,. \nn \\
\ea

Here I discuss the results of \cite{PRV09} and \cite{BP07}.
Previous work can be found in 
\cite{HKS95,HK98,BPP95,HK02,BPP02,KN02,KNP02,DB08,NYF09,MV04}
and recent reviews are in \cite{PR08,EdR09,JEG08,JN09}.

The hadronic four-point function 
$\Pi^{\mu\nu\rho\sigma}(q,k_1,k_3,k_2)$
is an extremely difficult object involving  many scales and  
no full first principle calculation of it has been reported yet --even in the simpler large numbers of colors $N_c$ limit of QCD.
Notice that we need it with momenta $k_1$, $k_2$ and $k_3$
 varying from 0 to $\infty$.
Unfortunately, unlike the hadronic vacuum polarization,
there is neither  a direct connection of $a^{\rm HLbL}$
 to a measurable quantity. Two lattice groups have started 
exploratory calculations \cite{HB06,RAK07}  but the 
final uncertainty that these calculations can reach is 
not clear yet.

Attending to a combined  large number of colors of QCD
 and chiral perturbation theory  (CHPT) counting, one can
distinguish four types of contributions \cite{EdR94}.
Notice that the CHPT counting is only for organization of the
contributions and refers to the lowest order term contributing
in each case. In fact,  Ref. \cite{PRV09} shows that there are 
chiral  enhancement factors that demand more than   
Nambu-Goldstone bosons in the CHPT expansion in the light-by-light
contribution to the muon anomaly. See more comments on this 
afterwards.

\vspace*{0.2cm}
The four different types of contributions are:
\begin{itemize}
\item Nambu-Goldstone boson exchanges contribution are ${\cal O}(N_c)$
and start  at ${\cal O}(p^6)$ in CHPT.
\item One-meson irreducible vertex contribution and 
non-Goldstone boson exchanges contribute  also at
${\cal O}(N_c)$  
but start contributing at ${\cal O} (p^8)$ in CHPT.
\item One-loop of Goldstone bosons contribution
 are ${\cal O}(1/N_c)$ and start at ${\cal O}(p^4)$ in CHPT.
\item One-loop of non-Goldstone boson contributions 
which are ${\cal O}(1/N_c)$  but start contributing  
at ${\cal O}(p^8)$ in CHPT.
\end{itemize}
Based on the counting above there are two full calculations
\cite{HKS95,HK98,HK02}  and \cite{BPP95,BPP02}. 
There is also a detailed study of the $\pi^0$
exchange contribution \cite{KN02} putting emphasis in obtaining
analytical expressions for this part.

Recently, two new calculations of the pion exchange
using  also the organization above have been made. In Ref.
\cite{DB08}, the pion pole term exchange is evaluated  
within an effective chiral model.
These authors  also study the box diagram
one-meson irreducible vertex contribution. 
The results are numerically very similar to the ones found in the 
literature as  can be seen in Table \ref{tab1}.
In Ref. \cite{NYF09}, the author uses a large $N_c$ model 
$\pi^0 \gamma^* \gamma^*$ form factor with the pion also of-shell. 
This has to be considered as a first step and more work 
has to be done in order to have the full light-by-light within this approach. 
In particular, it would be very interesting to calculate the 
contribution of one-meson irreducible vertex contribution within  
this model.

Using  operator product expansion (OPE) in QCD, 
the authors of \cite{MV04} pointed out a new short-distance
 constraint of the reduced full four-point Green function
\ba
\langle 0 | T \left[ 
V^\nu (k_1) V^\rho (k_3) V^\sigma (-(k_1+k_2+q)) \right]
| \gamma(q) \rangle  
\ea
when $q \to 0$  and in the special momenta configuration 
 $-k_1^2 \simeq -k_3^2 >> -(k_1+k_3)^2$ Euclidean and large.
In that kinematical region,
\ba
T \left[ V^\nu (k_1) V^\rho (k_3)\right]
\sim \frac{1}{\hat {k}^2} \, \varepsilon^{\nu\rho\alpha\beta}
\hat k_\alpha \, \left[\overline q \, {\hat Q}^2 \, \gamma_\beta 
\gamma_5 \, q \right] 
\ea
with $\hat k = (k_1-k_3)/2 \simeq k_1 \simeq - k_3$\, .
See also \cite{KPM04}. This short distance constraint was not 
explicitly imposed in previous to \cite{MV04} calculations.

\section{Leading in $1/N_c$  Results}

Using effective field theory techniques, the authors of
 \cite{KNP02} shown that leading large $N_c$ contribution 
to $a^{\rm HLbL}$  contains an enhanced term at low energy 
by  $\log^2 (M_\rho/m_\pi)$  where the rho mass $M_\rho$ acts as  an 
ultraviolet scale and the pion mass $m_\pi$ provides the infrared 
scale.  
\ba
a^{\rm HLbL}(\pi^0) = \left( \frac{\alpha}{\pi} \right)^3 \, N_c 
\frac{m^2 N_c}{48 \pi^2 f_\pi^2} \, \left[ \ln^2 \, \frac{M_\rho}{m_\pi}
+ {\cal O} \left( \ln \, \frac{M_\rho}{m_\pi} \right) + {\cal O} (1) \right] 
\ea

This leading logarithm is generated  by the Goldstone boson exchange
contributions and is fixed by the Wess--Zumino--Witten (WZW) 
vertex $\pi^0 \gamma \gamma$.  
In the chiral limit where quark
 masses are neglected and at large $N_c$, the coefficient of this 
double logarithm  is model  independent and has been calculated   
and shown to be  positive in \cite{KNP02}. 
All the calculations we discuss here 
agree with these leading behaviour  and its coefficient including the 
sign.  A global sign mistake in the $\pi^0$ exchange 
in \cite{HKS95,HK98,BPP95}
was found by \cite{KN02,KNP02} and confirmed by \cite{HK02,BPP02} 
and by others \cite{BCM02,RW02}. The subleading 
ultraviolet scale $\mu$-dependent terms \cite{KNP02}, namely, 
$\log(\mu/m_\pi)$ and a non-logarithmic term $\kappa(\mu)$, 
 are model dependent and calculations of them are implicit in the
results presented in  \cite{HKS95,HK98,BPP95,BPP02,MV04}. 
 In particular, $\kappa(\mu)$ contains the large $N_c$ 
contributions from  one-meson irreducible vertex 
and non-Goldstone boson exchanges.
In the next section we review the recent model calculations
of the full leading in the $1/N_c$ expansion contributions.

\subsection{Model Calculations}

The pseudo-scalar exchange  is the dominant numerical 
contribution  and was saturated in  
\cite{HKS95,HK98,BPP95,HK02,BPP02,KN02,DB08,NYF09} 
by Nambu-Goldstone boson's exchange. 
This contribution is depicted in Fig. 2
with $M \, = \, \pi^0,\eta,\eta^\prime$.
The relevant four-point function  was obtained 
in terms of the off-shell  $\pi^0 \gamma^*(k_1) \gamma^*(k_3)$
form factor ${\cal F}(k_1^2, k_3^2)$ and the off-shell
$\pi^0 \gamma^*(k_2) \gamma(q=0)$ form factor
${\cal F}(k_2^2, 0)$   modulating each one of the two 
 WZW $\pi^0 \gamma \gamma$  vertex. 
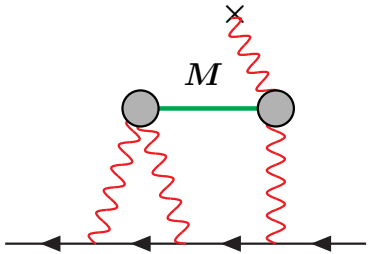
\begin{figure}
\label{pionexchange}
\begin{center}
\unitlength=1.5pt
\begin{picture}(80,80)
\SetScale{1.7}
\SetWidth{0.5}
\ArrowLine(20,0)(0,0)
\ArrowLine(40,0)(20,0)
\ArrowLine(60,0)(40,0)
\ArrowLine(80,0)(60,0)
\Text(58,58)[c]{\boldmath$\times$}
\SetColor{Red}
\Photon(20,0)(30,30){2}{6}
\Photon(40,0)(30,30){2}{6}
\Photon(60,0)(60,30){2}{6}
\Photon(60,30)(50,50){2}{5}
\SetColor{Green}
\SetWidth{1.}
\Line(30,30)(60,30)
\SetWidth{0.5}
\Text(50,40)[b]{\SetColor{Green}\boldmath$M$}
\SetColor{Black}
\GCirc(30,30){4}{0.7}
\GCirc(60,30){4}{0.7}
\end{picture}
\end{center}
\caption{A generic meson exchange contribution to the hadronic light-by-light
part of the muon $g-2$.}
\end{figure}
 
In all cases several short-distance QCD  constraints 
were imposed on these form-factors.
 In particular, they all have the correct QCD short-distance behaviour
\be
{\cal F}(Q^2, Q^2) \to \frac{A}{Q^2}
\hspace*{0.5cm} {\rm and} \hspace*{0.5cm} 
{\cal F}(Q^2, 0) \to \frac{B}{Q^2}
\ee
when $Q^2$ is Euclidean and large  and are  
in agreement with $\pi^0\gamma^*\gamma$ low-energy data
\footnote{See however the new measurement of the $\gamma \gamma^* \to \pi_0$ 
transition form factor by BaBar 
\cite{babar09} at energies between 4 and 40 GeV$^2$}.
\begin{table}
\begin{center}
\caption{Results for the $\pi^0$,
$\eta$ and $\eta'$ exchange contributions.
\label{tab1}}
{\begin{tabular}{c|cc}
 Reference &\multicolumn{2}{c}{ $10^{10} \times a$}\\
 & $\pi^0$ only &  $\pi^0$, $\eta$ and $\eta'$\\
\hline
\cite{HKS95,HK98,HK02}  & 5.7 & 8.3 $\pm$ 0.6 \\
\cite{BPP95,BPP02} & 5.6  & 8.5 $\pm$ 1.3 \\
\cite{KN02} with $h_2=0$ & 5.8 & 8.3 $\pm$ 1.2\\
\cite{KN02} with $h_2=-10$~GeV$^2$ & 6.3 & \\
\cite{DB08} & 6.3 $\sim$ 6.7 & \\
\cite{NYF09} & 7.2 & 9.9 $\pm$ 1.6 \\
\cite{MV04} &  7.65 &11.4$\pm$1.0  
\end{tabular}}
\end{center}
\vspace*{-0.5cm}
\end{table}
 They  differ slightly in shape  due to the different model
assumptions (VMD, ENJL, Large $N_c$, N$\chi$QM) 
but they produce small numerical differences always compatible 
within quoted uncertainty $\sim 1.3 \times 10^{-10}$ 
--see Table \ref{tab1}.

Within the models used in \cite{HKS95,HK98,BPP95,HK02,BPP02,KN02,DB08,NYF09}, 
to get the full contribution at leading  in $1/N_c$ 
one needs to add the one-meson irreducible vertex contribution and 
the non-Goldstone boson exchanges. In particular,  below some scale $\Lambda$,
the one-meson irreducible vertex contribution  was identified 
in \cite{BPP95,BPP02} with the ENJL quark box contribution with four dressed
photon legs. While to mimic the contribution of short-distance 
QCD quarks above $\Lambda$, a loop of bare massive heavy quark  with mass $\Lambda$
and QCD vertices was used.  The results are in Table \ref{quarkL} where one 
can see a very nice stability region when $\Lambda$ is in the interval
[0.7, 4.0] GeV.
Similar results for  the quark loop below $\Lambda$ 
were obtained in \cite{HKS95,HK98}  though these authors didn't discuss the 
short-distance long-distance  matching.
\begin{table}
\begin{center}
\caption{Sum of the short- and long-distance 
quark loop contributions  \cite{BPP95}
as a function of the matching scale $\Lambda$.
\label{quarkL}}{
\begin{tabular}{c|cccc}
$\Lambda$ [GeV] & 0.7 & 1.0 & 2.0 &4.0\\
\hline
\rule{0cm}{13pt} $10^{10} \times a^{\rm HLbL}$ & 2.2 &  2.0& 1.9& 2.0
\end{tabular}}
\end{center}
\vspace*{-0.5cm}
\end{table}

In \cite{BPP95,BPP02}, non-Goldstone boson exchanges were saturated by the 
hadrons appearing in the model, i.e. the lowest scalar and pseudo-vector
hadrons. Both states
in nonet-symmetry --this symmetry is exact in the large $N_c$ limit. 
Within the ENJL model, the one-meson irreducible vertex contribution  
 is related trough Ward  identities  to the scalar exchange 
which we discuss below and {\it both}  have to be included \cite{BPP95,BPP02}.
 The result  of the scalar exchange obtained in \cite{BPP95} is 
\be
\label{scalar}
a^{\rm HLbL}(\rm Scalar) = -(0.7 \pm 0.2 ) \times 10^{-10} \, .
\ee
The scalar exchange was not included in \cite{HKS95,HK98,HK02,KN02}.
The result of the axial-vector exchanges in \cite{HKS95,HK98,HK02}
and \cite{BPP95,BPP02} can be found in Table \ref{tab3}.
\begin{table}
\begin{center}
\caption{Results for the axial-vector exchange contributions from 
\cite{HKS95,HK98,HK02} and \cite{BPP95,BPP02}.\label{tab3}}{
\begin{tabular}{c|c}
 References
 & $10^{10} \times a^{\rm HLbL}$\\
\hline
 \cite{HKS95,HK98,HK02}  & 0.17 $\pm$ 0.10  \\
\cite{BPP95,BPP02}  & 0.25 $\pm$ 0.10 
\end{tabular}}
\end{center}
\vspace*{-0.5cm}
\end{table}

 Melnikov and Vainshtein used a model that saturates
the hadronic four-point function in (\ref{four}) at
leading order in the $1/N_c$ expansion by the exchange 
of the Nambu-Goldstone $\pi^0, \eta, \eta'$ and 
the lowest axial-vector $f_1$ states.
 In that model, the new OPE constraint of the reduced four-point
function found in \cite{MV04} mentioned above, 
forces the $\pi^0 \gamma^*(q) \gamma(p_3=0)$ 
vertex to be point-like rather than including a ${\cal F}(q^2,0)$ 
form factor. 
\begin{figure}
\begin{center}
\unitlength=1.5pt
\begin{picture}(80,80)
\SetScale{1.7}
\SetWidth{0.5}
\ArrowLine(20,0)(0,0)
\ArrowLine(40,0)(20,0)
\ArrowLine(60,0)(40,0)
\ArrowLine(80,0)(60,0)
\Text(58,70)[c]{\boldmath$\times$}
\SetColor{Red}
\Photon(20,0)(30,30){2}{6}
\Photon(40,0)(30,30){2}{6}
\Photon(60,0)(60,30){2}{6}
\Photon(60,30)(50,60){2}{6}
\SetColor{Blue}
\SetWidth{1.}
\Line(30,30)(60,30)
\SetWidth{0.5}
\Text(50,40)[b]{\SetColor{Green}\boldmath$\pi^0,\eta,\eta'$}
\SetColor{Black}
\GCirc(30,30){4}{0.7}
\Vertex(60,30){2}
\SetColor{Red}
\SetWidth{1.0}
\LongArrow(75,50)(62,33)
\Text(85,60)[l]{\SetColor{Green}\boldmath$Point-Like$}
\end{picture}
\end{center}
\caption{Goldstone boson exchange in the model in \cite{MV04}
contributing to the hadronic light-by-light.}
\label{pionexchangeMV}
\end{figure}
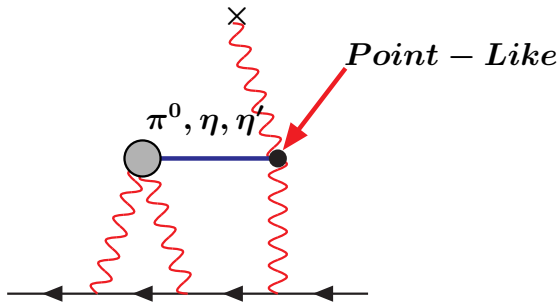
There are also OPE constraints for
other momenta regions \cite{NSV84} which are not satisfied by the model
in \cite{MV04} though they argued that this made
 only a small numerical difference of the order of $0.05 \times 10^{-10}$. 
 In fact, 
within the large $N_c$ framework, it has been shown \cite{BGL03}
that in general for other than two-point functions, to satisfy fully
the QCD short-distance properties requires
the inclusion of an infinite number of narrow states.

\section{Next-to-leading in $1/N_c$  Results}

 For the next-to-leading in $1/N_c$ contributions to
the $a^{\rm HLbL}$ there is no model independent result at present
 and is possibly the most difficult component.
 Charged pion  and kaon loops saturated this contribution
 in \cite{HKS95,HK98,BPP95,HK02,BPP02}.  To dress the photon 
interacting  with pions, a particular Hidden Gauge Symmetry (HGS) 
model was used in  \cite{HKS95,HK98,HK02} while  a full VMD was used 
in \cite{BPP95,BPP02}. 
The results obtained are $-(0.45\pm 0.85) \times 10^{-10}$ in \cite{HKS95}
and $-(1.9\pm0.5) \times 10^{-10}$ in \cite{BPP95} while using a point-like
vertex one gets $- 4.6 \times 10^{-10}$.
Both models satisfy the known constraints  
though start differing at  ${\cal O}(p^6)$ in CHPT.
 Some studies of the cut-off dependence 
of the pion loop using the full VMD model 
was done  in \cite{BPP95}  and showed  that  their  final number comes 
from fairly low energies where the model dependence should be smaller.
The authors of \cite{MV04} analyzed the model
used in \cite{HKS95,HK98} and  showed that there is a large 
cancellation between the first three terms of an expansion in powers of 
$(m_\pi/M_\rho)^2$ and  with large higher order corrections
when expanded in CHPT orders but the same applies to the $\pi^0$
exchange as can be seen from Table 6 in the first reference in \cite{BP07} 
by comparing the WZW  column with the others.
 The authors of \cite{MV04} took 
$(0\pm 1) \times 10^{-10}$ as  a guess estimate of the total NLO in $1/N_c$
contribution. This seems too simply and  certainly with 
underestimated uncertainty.

\section{Comparing Different Calculations}

The comparison of individual contributions 
in \cite{HKS95,HK98,BPP95,HK02,BPP02,KN02,DB08,NYF09,MV04}   
has to be done with care because they come from different model assumptions
to construct the full relevant four-point function. 
In fact,  the authors of \cite{DB08} have shown that their
constituent quark loop  provides the correct asymptotics
and in particular the new OPE found in \cite{MV04}. 
It has more sense to compare  results for $a^{\rm HLbL}$ either  at 
leading order or at next-to-leading order in the $1/N_c$ expansion. 

The results for the final hadronic light-by-light 
contribution to $a^{\rm HLbL}$ quoted in \cite{HKS95,HK98,BPP95,HK02,BPP02,MV04}
 are in Table \ref{tab4}.
The apparent agreement between \cite{HKS95,HK98,HK02} and \cite{BPP95,BPP02}
hides non-negligible differences which numerically almost compensate
between the quark-loop and charged pion and \cite{MV04} are in Table \ref{tab4}.
 Notice also that \cite{HKS95,HK98,HK02} didn't include the scalar exchange. 
\begin{table}
\begin{center}
\caption{Results for the full hadronic light-by-light
contribution to $a^{\rm HLbL}$.\label{tab4}}{
\begin{tabular}{c|c}
 Full Hadronic Light-by-Light
 & $10^{10} \times a_\mu$\\
\hline
 \cite{HKS95,HK98,HK02}  & 8.9$\pm$ 1.7  \\
\cite{BPP95,BPP02}  & 8.9 $\pm$ 3.2  \\
\cite{MV04}         & 13.6 $\pm$ 2.5 
\end{tabular}}
\end{center}
\vspace*{-0.5cm}
\end{table}
 Comparing the results of \cite{BPP95,BPP02} and \cite{MV04},
as discussed above, we have found several differences of order
$1.5 \times 10^{-10}$  which are not related to the new short-distance
constraint used in \cite{MV04}.  The different  axial-vector mass
mixing accounts for $-1.5 \times 10^{-10}$, the absence of the
scalar exchange in \cite{MV04} accounts for $-0.7 \times 10^{-10}$
and the use of a vanishing  NLO in $1/N_c$ contribution
in \cite{MV04} accounts for $-1.9 \times 10^{-10}$. These model
dependent differences add up to $-4.1 \times 10^{-10}$ out of the
final $-5.3 \times 10^{-10}$ difference between \cite{BPP95,BPP02}
and \cite{MV04}. Clearly, the new OPE constraint  used
in \cite{MV04} accounts only for a small part 
of the large numerical final difference.

\section{Conclusions and Prospects}

To give a result at present for the hadronic 
light--by--light contribution to the muon anomalous magnetic moment,  
the authors of \cite{PRV09},  from  the above considerations, 
 concluded that it is fair to proceed as follows

\noi
{\it Contribution to $a^{\rm HLbL}$ from $\pi^0$, $\eta$ and $\eta'$ exchanges}\\[1mm]
Because of the effect of the OPE constraint discussed above, we suggested to take as central value the result of Ref.~\cite{MV04} with, however,  the largest error quoted in Refs.~\cite{BPP95,BPP02}:
\be
a^{\rm HLbL}(\pi\,,\eta\,,\eta')=(11.4\pm 1.3)\times 10^{-10}\,.
\ee
Recall that this central value is quite close to the one in the ENJL 
model when the short--distance quark loop contribution  is added there.

\noi
{\it Contribution to $a^{\rm HLbL}$ from pseudo-vector exchanges}\\[1mm]
The analysis made in Ref.~\cite{MV04} suggests that the errors in the first and second entries of Table~2 are likely to be underestimates. Raising their $\pm 0.10$ errors  to $\pm 1$ puts the three numbers in agreement within one sigma. We suggested then as the best estimate at present
\be
a^{\rm HLbL}(\rm{pseudo-vectors})=(1.5\pm 1)\times 10^{-10}\,.
\ee

\noi
{\it Contribution to $a^{\rm HLbL}$ from scalar exchanges}\\[1mm]
The ENJL--model should give a good estimate for these contributions. 
We kept, therefore, the result of Ref.~\cite{BPP95,BPP02} with, 
however, a larger error which covers the effect of other
 unaccounted meson  exchanges,
\be
a^{\rm HLbL}(\rm{scalars})=-(0.7\pm 0.7)\times 10^{-10}\,.
\ee

\noi
{\it Contribution to $a^{\rm HLbL}$ from a dressed pion loop}\\[1mm]
Because of the instability of the results for the charged pion loop and 
unaccounted loops of other mesons, we suggested using the central 
value of the ENJL result  but wit a larger error:
\be
a^{\rm HLbL}(\pi{\rm -dressed~loop})=-(1.9\pm 1.9)\times 10^{-10}\,.
\ee

{}From these considerations, adding the errors in quadrature,
 as well as the small charm contribution $0.23 \times 10^{-10}$, we get
\be
a^{\rm HLbL}=(10.5\pm 2.6)\times 10^{-10}\,,
\ee
as our final estimate. 

The proposed new $g_\mu-2$ experiment 
accuracy goal of $1.6 \times 10^{-10}$ calls
for a considerable improvement in the present
calculations. The use of further theoretical and experimental constraints could
 result in reaching such accuracy soon enough. In particular, imposing as many 
as possible short-distance QCD constraints 
\cite{HKS95,HK98,BPP95,HK02,BPP02,KN02,NYF09} 
has result in a better understanding  of the numerically dominant $\pi^0$ exchange. 
At present, none of the  light-by-light hadronic parametrization satisfy fully 
all short distance QCD constraints. 
In particular, this requires the  inclusion  of
 infinite number of narrow states for other 
than two-point functions  and two-point functions with soft insertions \cite{BGL03}.
A numerical dominance of certain momenta configuration
can help to minimize  the effects of short distance QCD constraints
 not satisfied, as in the model in \cite{MV04}.

More experimental information on the decays $\pi^0 \to \gamma \gamma^*$,
$\pi^0 \to \gamma^* \gamma^*$ and $\pi^0 \to e^+ e^-$ 
(with radiative 
corrections included \cite{RW02,KKN06,KM09}) 
can also help to confirm some of the neutral pion exchange results.
A better understanding of other smaller 
contributions but with comparable uncertainties needs both more
 theoretical work and  experimental information.
This refers in particular to  pseudo-vector exchanges. 
Experimental data on radiative decays and two-photon 
production of these and  other C-even resonances can be useful in that respect. 

New approaches to the pion dressed loop contribution, together with experimental
information on the vertex $\pi^+\pi^-\gamma^*\gamma^*$  in the intermediate 
energy region ($0.5-1.5$ GeV) would 
also be very welcome. Measurements of two-photon processes like 
$e^+e^- \to e^+e^-\pi^+\pi^-$  can be useful to give information on that 
vertex and again could reduce the model dependence.
 The two-gamma physics program low energy facilities like the experiment
KLOE-2 at DA$\Phi$NE will be very useful
 and well suited  in the processes mentioned above which
 information can help  to decrease the present model dependence 
of $a^{\rm HLbL}_\mu$.

\section{Acknowledgments}

It is a pleasure to thank Hans Bijnens,
Elisabetta Pallante, Eduardo de Rafael and Arkady Vainshtein
for enjoyable collaborations and discussions with Andreas Nyffeler
on  the different topics    discussed here.
This work has been supported in part  by MICINN,
 Spain and FEDER, European Commission (EC) Grant No. FPA2006-05294, 
by the Spanish Consolider-Ingenio 2010 Programme CPAN 
Grant No. CSD2007-00042,  by Junta de Andaluc\'{\i}a Grants No. 
 P05-FQM 347 and P07-FQM 03048 and  by the EC RTN FLAVIAnet
 Contract No. MRTN-CT-2006-035482.

\section{Bibliography}




\end{document}